\documentclass[10pt]{artikel3}
\usepackage{amsfonts,amsmath,amssymb,hyperref}
\usepackage{pslatex}

\usepackage[dvips]{feynmp}

\usepackage{epsfig}
\usepackage{dsfont}

\usepackage{float,braket}
\usepackage{subfigure}
\usepackage{amsmath}
\usepackage{amsfonts}
\usepackage{amssymb}

\usepackage{graphicx,wrapfig,color}

\newcommand{\MSbar}{\overline{\mbox{MS}}}

\newcommand{\p}{\partial}
\newcommand{\oc}{\overline{c}}

\newcommand{\s}{\sigma}
\newcommand{\omu}{\overline{\mu}}
\newcommand{\lms}{\Lambda_{\overline{\mbox{\tiny{MS}}}}}

\renewcommand{\d}{\ensuremath{\mathrm{d}}}

\renewcommand{\d}{\ensuremath{\mathrm{d}}}

\newcommand{\gf}{\ensuremath{\mathrm{gf}}}
\newcommand{\YM}{\ensuremath{\mathrm{YM}}}

\newcommand{\GZ}{\ensuremath{\mathrm{GZ}}}

\definecolor{Rood}{rgb}{1, 0, 0} 

\begin{document}
\title{\noindent {\bf On bounds and boundary conditions in the continuum Landau gauge}}
\author{D.~Dudal$^{a,b}$\thanks{david.dudal@kuleuven-kulak.be},\; M.S.~Guimaraes$^{c}$\thanks{msguimaraes@uerj.br},\;I.F.~Justo$^{b,c}$\thanks{igorfjusto@gmail.com
},\; S.P.~Sorella$^{c}$\thanks{sorella@uerj.br}\\\\
{\small  \textnormal{$^{a}$ KU Leuven Campus Kortrijk - KULAK, Department of Physics, Etienne Sabbelaan 53, 8500 Kortrijk, Belgium}}
\\
{\small \textnormal{$^{b}$ Ghent University, Department of Physics and Astronomy, Krijgslaan 281-S9, 9000 Gent, Belgium}}\\
\small \textnormal{$^{c}$ Departamento de F\'{\i }sica Te\'{o}rica, Instituto de F\'{\i }sica, UERJ - Universidade do Estado do Rio de Janeiro}
 \normalsize}

\date{}
\maketitle
\begin{abstract}
In this note, we consider the Landau gauge in the continuum formulation. Our purposes are twofold. Firstly, we try to work out the consequences of the recently derived Cucchieri-Mendes bounds on the inverse Faddeev operator at the level of the path integral quantization. Secondly, we give an explicit (renormalizable) prescription to implement the so-called Landau $B$-gauges as introduced by Maas.
\end{abstract}

\setcounter{page}{1}

\section{Introduction}
The Faddeev-Popov operator $\mathcal{M}\equiv \mathcal{M}^{ab}[A]=-\p_\mu D_\mu^{ab}$, with $D_\mu^{ab}=\delta^{ab}\p_\mu -gf^{abc}A_\mu^c$ the covariant derivative in the adjoint representation, plays an important role when discussing geometric features of the Landau gauge fixing $\p_\mu A_\mu^a=0$ for non-Abelian gauge theories. The set of all $A_\mu\equiv A_\mu^a$ that fulfill
\begin{equation}\label{g1}
  \p_\mu A_\mu=0 ~\wedge~ \mathcal{M}\geq 0
\end{equation}
defines the so-called Gribov region $\Omega$. At its boundary $\p\Omega$, the Faddeev-Popov operator becomes zero. The definition \eqref{g1} makes sense, since $\mathcal{M}$ is a Hermitian operator, thus having real eigenvalues. It is then a simple matter to see that $A_\mu^a\in\Omega$ is a necessary and sufficient condition to ensure that $A_\mu$ has no \emph{infinitesimally} gauge equivalent field configurations also subject to the Landau gauge. The Gribov gauge fixing ambiguity at the infinitesimal level is thus overcome if the set of admissible gauge fields in the path integral are those belonging to $\Omega$. A few interesting properties of the region $\Omega$ have been established: it is convex, bounded in every direction (it is a kind of multidimensional ellipsoid) and most importantly, each gauge orbit intersects $\Omega$ at least once \cite{Zwanziger:1982na,Dell'Antonio:1991xt}.

The conditions \eqref{g1} are nothing else than those for identifying a local minimum of the functional \cite{Semenov,Vandersickel:2012tz}
\begin{equation}\label{g1b}
  \mathcal{R}[A]=\min_{U\in \text{SU}(N)}\int \d^4 x A_\mu^U A_\mu^U.
\end{equation}

If we think in terms of the action of Landau gauge-fixed Yang-Mills theory,
\begin{eqnarray}\label{g2}
S_{GF+FP} &=& \frac{1}{4}\int \d^d x\; F^a_{\mu\nu} F^a_{\mu\nu}+\int \d^d x\,\left( b^a \p_\mu A_\mu^a +\overline c^a \p_\mu D_\mu^{ab} c^b \right)\,,
\end{eqnarray}
the inverse Faddeev-Popov operator can be identified with the ghost propagator $\mathcal{G}^{ab}(k,A)$ in the presence of an external gauge field. This observation is the crux of the original recipe \cite{Gribov:1977wm} of Gribov to work out a practical version of the restriction to $\Omega$: only gauge fields with $\sigma(k,A)\leq 1$ must be taken into account in the path integral, where $G(k,A)=\frac{1}{N^2-1}\mathcal{G}^{aa}(k,A)=\frac{1}{k^2}\frac{1}{1-\sigma(k,A)}$. This is known as the no-pole condition and it was translated into an effective action at tree level by Gribov. Using an alternative derivation, Zwanziger was able to generalize the Gribov effective action to all orders \cite{Zwanziger:1989mf,Zwanziger:1992qr}. The end point was the following (Gribov-Zwanziger) action
\begin{eqnarray}\label{g2b}
S_\GZ &=& \frac{1}{4}\int \d^d x\; F^a_{\mu\nu} F^a_{\mu\nu}+\int \d^d x\,\left( b^a \p_\mu A_\mu^a +\overline c^a \p_\mu D_\mu^{ab} c^b \right)\nonumber\\&&+\int \d^d x \left( \overline{\varphi }_{\mu }^{ac} \partial _{\nu} D_\nu^{am}\varphi _{\mu
}^{mc} -\overline{\omega }_{\mu }^{ac} \partial _{\nu } D_\nu^{am} \omega _{\mu }^{mc}
  -g\left( \partial _{\nu }\overline{\omega }_{\mu}^{ac}\right) f^{abm}\left( D_{\nu }c\right) ^{b}\varphi _{\mu
}^{mc}\right)\nonumber\\
&&+\int\d^d x\left( -\gamma ^{2}gf^{abc}A_{\mu }^{a}(\varphi _{\mu }^{bc}+\overline{\varphi }_{\mu }^{bc}) - d\left(N^{2}-1\right) \gamma^{4} \right) \,.
\end{eqnarray}
The fields $\varphi_{\mu }^{ac}$, $\overline{\varphi }_{\mu }^{ac}$ are a set of bosonic fields each with $d(N^2-1)^2$ components, the fields $\omega_{\mu }^{ac}$, $\overline{\omega}_{\mu }^{ac}$ are ghost fields. The Gribov mass parameter $\gamma^2$ is not free, but it is determined in a self-consistent way through the gap equation $\left.\frac{\p E}{\p \gamma^2}\right|_{\gamma^2\neq0}=0$, where $E$ stands for the vacuum energy of the theory \cite{Gribov:1977wm,Zwanziger:1989mf,Zwanziger:1992qr}. In more recent years, it was shown that nontrivial vacuum effects can further alter the action, leading to the so-called Refined Gribov-Zwanziger (RGZ action)  \cite{Dudal:2007cw,Dudal:2008sp,Dudal:2011gd,Gracey:2010cg}. It is also worthwhile to mention that recently, progress has been made to generalize the Gribov-Zwanziger construction to the Maximal Abelian gauge \cite{Capri:2008vk,Guimaraes:2011sf,Gongyo:2013rua,Pereira:2014apa,Capri:2013vka}. Also studies of Gribov copies in curved space times were considered \cite{Canfora:2011xd,Canfora:2013mrh}, next to theories including Higgs scalar fields \cite{Capri:2012ah,Capri:2013oja} or  supersymmetric gauge theories \cite{Capri:2014xea,Capri:2014tta}.

An alternative way to cope with the Gribov ambiguity, based on sampling over all copies, has been put forward in \cite{Serreau:2012cg}. The net result was the \emph{massive} Curci-Ferrari action \cite{Curci:1976bt,Curci:1976kh}, where the presence/value of the mass is, to our knowledge, a somewhat delicate issue related to taking a set of limits, see also \cite{vonSmekal:2008en} for similar questions. Nonetheless, a decent effective description of two- and higher point functions was feasible \cite{Tissier:2010ts,Reinosa:2013twa,Pelaez:2014mxa}.

In the Gribov-Zwanziger approach, we are implicitly assigning an equal weight to all further copies inside $\Omega$ and sample over those. In the literature, this is sometimes referred to as the minimal Landau gauge. Such sampling might be ill-posed, if the number of copies inside $\Omega$, $\mathcal{N}[A]$, depends on the gauge field, but it is the best we can do analytically. Almost nothing is known about $\mathcal{N}[A]$. Let us refer to \cite{Maas:2011se} for a more elaborate discussion on the potential pitfalls of extending the Landau gauge beyond its perturbative definition.

It is perhaps also worthwhile to remind that $\Omega$ itself is not free from residual gauge copies. This was discussed in mathematical detail in \cite{Semenov}, see also \cite{vanBaal:1991zw}. Ideally, one would have to resort to restricting the path integral to a region free of Gribov copies, defining an absolute Landau gauge. This so-called fundamental modular region (FMR) in the Landau gauge would correspond to the region $\Lambda\subset\Omega$ built from the absolute (in contrast to the local) minima of the functional $\mathcal{R}[A]$, \eqref{g1b}. Unfortunately, there are no simple criteria which a generic $A_\mu^a$ should fulfill to belong to the FMR $\Lambda$. Not much is known about $\Lambda$, apart from the fact that it is also convex \cite{Semenov} and evidently, it is also bounded in every direction. Even numerically, it is extremely hard to locate (all) absolute minima of a functional in a complicated space. This is immediately the main reason why the Gribov-Zwanziger restriction to $\Omega$ is mostly used in practical applications.

Even $\Lambda$ itself is not free of copies, its boundary $\p\Lambda$ can display degenerate absolute minima as shown in \cite{Zwanziger:1993dh}. It is also known that $\Omega$ and $\Lambda$ share part of their boundary \cite{Zwanziger:1993dh}. In e.g.~\cite{Zwanziger:2003cf}, it was conjectured and analytically motivated that $\braket{\ldots}_\Omega=\braket{\ldots}_\Lambda$. Until now, this remains however a conjecture, and not all lattice studies comply with it \cite{Bornyakov:2013ysa}. In any case, interpreting lattice results with regard to the (continuum, thus infinite-volume) conjecture should be done with care, since the infinite volume is never reached.

The presence of copies inside the Gribov region $\Omega$ opens the door for alternative completions of the Landau gauge, see also \cite{Maas:2013vd} for a discussion. For example, one could pick, per gauge configuration, as the relevant contribution to the path integral a copy strictly inside $\Omega$ and not the one on the boundary $\p\Omega$ \cite{Silva:2004bv}. Gauge dependent quantities will in general be dependent on the specific copy selection\footnote{Looking forward to Section 3, a different selection of copies could lead to a different value for the parameter $B$. Vice versa, a priori picking $B$ could then correspond to a different selection (sampling) over Gribov copies. }. Though, in principle, when evaluating gauge invariant quantities, it should be of no consequence whatever copy is used. Though, such selection procedure is at the practical level only feasible in the lattice context, and probably even only in at relatively small volumes, for which there is relative control on the set of Gribov copies.

In functional formalisms, with typical proponent the Dyson-Schwinger equations \cite{Alkofer:2000wg,Roberts:1994dr,Binosi:2009qm,Boucaud:2011ug}, it was discussed in \cite{Zwanziger:2001kw,Huber:2009tx} that the form/solutions of the equations should not change by cutting off the region of integration at the boundary $\p\Omega$.

In the following Section, an analytical argument will be provided that ---at least in the employed continuum setting--- the path integral over the Gribov region $\Omega$ is still dominated by the integration over its boundary $\p\Omega$ when making use of the Cucchieri-Mendes bound on the ghost propagator \cite{Cucchieri:2013nja,Cucchieri:2013xka}. In the third Section, we focus on answering a question, originally posed by Maas in \cite{Maas:2009se,Maas:2010wb} and improved upon in \cite{Maas:2011se}, concerning the implementation at the functional level of a boundary condition on the (inverse) ghost propagator. Besides that, also some quantum aspects of the proposed action, including its renormalizability, are handled in the fourth Section. We end with a more concise discussion of the obtained results.

\section{The Cucchieri-Mendes bound on the ghost propagator and its implications}
\begin{figure}[t]
  \centering
  \scalebox{0.5}{\includegraphics{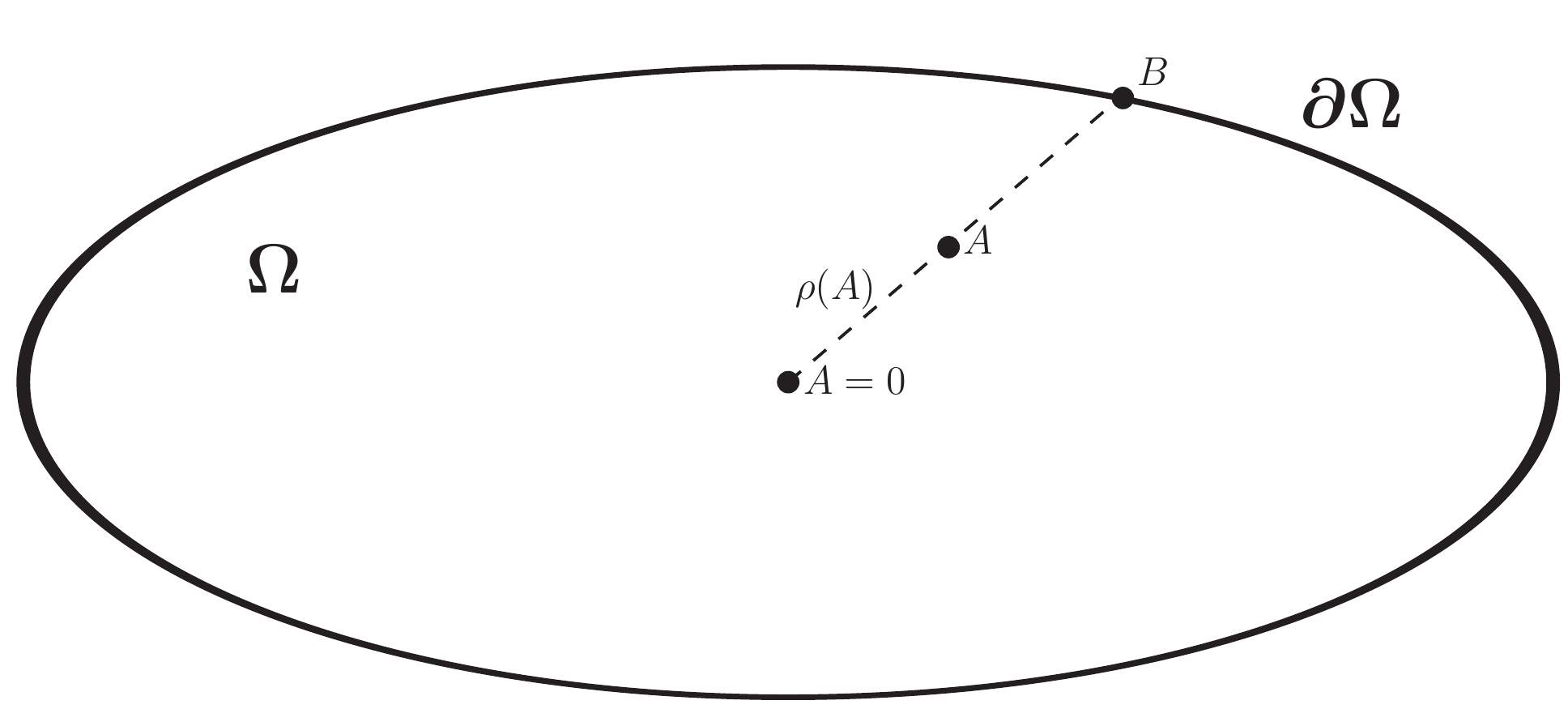}}
  \caption{Pictoral representation of the Gribov region $\Omega$ and the distance $\rho(A)$.}\label{figure}
\end{figure}

In \cite{Cucchieri:2013nja}, it was derived how the horizon function, defined via
\begin{equation}\label{g2c}
  h(A)=\frac{1}{Vd(N^2-1)}g^2 \int \d^dx \d^dy f^{abc}A_\mu(x) \left[\mathcal{M}^{-1}\right]^{ad}f^{dec}A_\mu(y)\,,\quad [V=\text{space-time volume}]\,,
\end{equation}
obeys the following inequality,
\begin{equation}\label{cm1}
  h(A)\leq \rho(A)\,,
\end{equation}
where $1-\rho(A)$ is the normalized distance of the gauge configuration $A$ to $\p\Omega$, or, equivalently, $\rho(A)$ is the normalized distance to $0\in \Omega$. This distance can be defined on basis of the convexity of $\Omega$ \cite{Zwanziger:1982na}, in the sense that for each $A\in\Omega$, we can consider in $\Omega$ the ray connecting $A_\mu$ with $0$, calling $B_\mu$ the intersection of that ray with $\p\Omega$. Doing so, $\rho(A)$ corresponds to the scaling factor in the relation $A_\mu= \rho(A) B_\mu$. In more mathematical terms \cite{Cucchieri:2013nja}, introducing the $\mathcal{L}_2$ norm for vector valued SU($N$) matrices,  $||A||^2=\frac{1}{dV(N^2-1)}\int \d^dx \text{Tr}\left[A_\mu A_\mu^\dagger\right]$, then $\rho(A)= \frac{||A ||}{||B||}$.  Notice that by construction $\rho(A)\leq 1$. Figure \ref{figure} displays the situation in a pictorial manner.

In \cite{Gomez:2009tj,Capri:2012wx}, it was shown, order by order in a perturbative expansion, that the horizon function $h(A)$ corresponds to the Gribov ghost form factor $\sigma(k,A)$ at zero momentum. Imposing $\sigma(0,A)\leq 1$ (while keeping in mind that for the a posteriori constructed integration measure, $\braket{\sigma(k,A)}$ is a decreasing function in $k$) is thus equivalent to imposing $h(A)\leq 1$. In the thermodynamic limit \cite{Gribov:1977wm,Zwanziger:1989mf}, the inequality becomes an equality, see also \cite{Vandersickel:2012tz}, and using this we arrive at the horizon condition
\begin{equation}\label{hor}
  \braket{h(A)}=1 \Leftrightarrow \braket{\sigma(0,A)}^{1PI}=1\,,
\end{equation}
where we added the superscript $1PI$ to make clear that reference is made to the 1-particle-irreducible part of the ghost self-energy \cite{Capri:2012wx}. Evidently, the measure underpinning $\braket{\ldots}$ in \eqref{hor} is not the Faddeev-Popov one, since $\sigma(0)\equiv \braket{\sigma(0,A)}^{1PI}\neq 1$ using standard perturbation theory. Let us also remind here that $\sigma(0)=1$ leads to a more than quadratically IR divergent ghost propagator, a property first pointed out by Gribov \cite{Gribov:1977wm}, later on proven to all orders using a Ward identity argument in \cite{Zwanziger:1992qr} by Zwanziger. Naively, the Gribov-Zwanziger restriction boils down to taking into account only gauge configurations living on the Gribov horizon $\p\Omega$, where the ghost propagator diverges more strongly due to the presence of the zero modes.

Though, the observation of an IR enhanced ghost propagator is in disagreement with contemporary lattice simulations\footnote{The (minimal) Landau gauge can be numerically implemented using a minimization algorithm for $||A||^2$, as local minima are found for transverse gauge fields (vanishing 1st derivative) that belong to the Gribov region; the Faddeev-Popov operator indeed corresponds to the 2nd derivative which needs to be positive.} \cite{lattice1,lattice2,lattice3,lattice4,lattice5,lattice5b,lattice6,lattice7,lattice8,lattice9,Sternbeck:2012mf}, a fact later recognized in the Gribov-Zwanziger theory to be caused by a vacuum instability \cite{Dudal:2007cw,Dudal:2008sp,Dudal:2011gd,Gracey:2010cg} that further changes the effective action in a dynamical fashion. Similar results were found using other approaches \cite{Tissier:2010ts,Aguilar:2004sw,Aguilar:2008xm,Fischer:2008uz,Boucaud:2008ky}. For a nonextensive list of recent additions to these research programs, let us refer to \cite{Aguilar:2011yb,Aguilar:2013vaa,Binosi:2014kka,Aguilar:2014tka,Pennington:2011xs,Huber:2012kd,Huber:2014tva,Blum:2014gna,Cyrol:2014kca,Eichmann:2014xya,Boucaud:2011eh,Dudal:2012zx,Cucchieri:2012cb}.

Given the a priori sharper Cucchieri-Mendes bound \eqref{cm1}, the possibility seems to be lurking around the corner for a geometrical reason behind a non-enhanced ghost in the infrared, given that one is actually integrating over a \emph{smaller} subregion of the Gribov region $\Omega$, since for many gauge configurations it will hold that $\rho(A)<1$. Though, depending on the number of configurations that do have $\rho(A)=1$, ghost enhancement would still be allowed on mere \emph{classical/geometrical grounds}, but not realized explicitly because of another reason, in particular the dynamical refinement as discussed in \cite{Dudal:2007cw,Dudal:2008sp,Dudal:2011gd,Gracey:2010cg}, which would be a reason on \emph{quantum grounds}. We refer to \cite{Cucchieri:2013nja} for many more details, including lattice experiments and related discrete versions of the bound.

In the remainder of this Section, we come to our first result: directly working in the continuum, we shall show that the path integral over the Gribov region $\Omega$ is still dominated by the integration over the horizon $\p\Omega$. Indeed, consider the following path integral
\begin{equation}\label{p1}
  \int [\mathcal{D} A] \theta[\rho(A)-h(A)] \theta[1-\rho(A)]e^{-S_{FP}[A]}=\int_\Omega [\mathcal{D}A] \theta[\rho(A)-h(A)] e^{-S_{FP}[A]}=(\ast)
\end{equation}
which encompasses the Cucchieri-Mendes bound \eqref{cm1} via the first $\theta$-function, next to the restriction to the Gribov region via the second $\theta$-function. Now given that the region $\Omega$ is for sure star-convex, it is natural to use ``polar coordinates'' to rewrite the integral as an integral over the boundary $\Omega$ times a ``normalized radial'' integral\footnote{Notice that the norm of the $B$-fields is part of the $[\d B]$ measure since we normalized $\rho$ onto $[0,1]$. The $B$ are thus genuine gauge fields living on $\p\Omega$.};
\begin{equation}\label{p2}
(\ast)= \int_{\p\Omega} [\mathcal{D}B]  \int_0^{1} \d\rho \rho^N \theta[\rho-h(\rho B)] e^{-S_{FP}[\rho B]}\,,
\end{equation}
where $N=\dim(\Omega)-1=\infty$. We first rewrite the $B$-integral via the identification
\begin{equation}\label{p2b}
  \int_{\p\Omega} [\mathcal{D}B](\ldots)=\int [\mathcal{D}B] \delta(h(B)-1)(\ldots)\,.
\end{equation}
The latter integral can be interpreted as defining a microcanonical average. Making use of the thermodynamic limit \cite{Dudal:2009xh,Zwanziger:1989mf}, the former averages equal those generated by a Boltzmann distribution (see also later for some more details), defined by
\begin{equation}\label{p2c}
  \int_{\p\Omega} [\mathcal{D}B](\ldots)=\int [\mathcal{D}B] \delta(h(B)-1)(\ldots)\to\int [\mathcal{D} B] e^{\gamma^4\int \d^4x h(B)}(\ldots)\,,
\end{equation}
with $\gamma^4$ the critical (Gribov) parameter determined by the self-consistency equation
\begin{equation}\label{p2d}
  \braket{h}=1\,,
\end{equation}
where the measure adopted is the one corresponding to the r.h.s.~of \eqref{p2c}. We thus get for the partition function
\begin{equation}\label{p2e}
(\ast)= \int[\mathcal{D} B]  \int_0^{1} \d\rho \rho^N \theta[\rho-h(\rho B)] e^{-S_{FP}[\rho B]+\gamma^4\int \d^4x h(B)}\,.
\end{equation}
Then, we use the Laplace approximation rule for an integral of the form
\begin{equation}\label{p3}
\lim_{N\to\infty}  \int_0^1 \d xx^N f(x)=\lim_{N\to\infty}  \int_0^1 \d x e^{N\ln x} f(x)\propto \lim_{N\to\infty} e^{N\ln x_0} f(x_0)\,,
\end{equation}
where $x_0=1$ is the maximum of  $\ln x$ over the interval $[0,1]$. Thus, it is clear the $\rho$-integral in \eqref{p2e} will be reduced to its value at $\rho=1$ (i.e.~on the horizon). We find for our partition function
\begin{equation}\label{p4}
(\ast)= \int[\mathcal{D} B]  \theta[1-h(B)] e^{-S_{FP}[B]+\gamma^4\int \d^4x h(B)}\,.
\end{equation}
In the latter expression, we recognize almost the standard Gribov-Zwanziger action. The difference lies in the presence of the $\theta[1-h(B)]$, but this is actually an obsolete factor. Indeed, thanks to the thermodynamic limit/infinite-dimensionality of the configuration space \cite{Vandersickel:2012tz} the $\theta$-function collapses into a $\delta$-function\footnote{The usual argument employed is the analogy with the sphere in $D$ dimensions. For growing $D$, the volume of the sphere gets more and more concentrated on its boundary.}, and we can repeat the microcanonical~$\to$~Boltzmann averaging. Though, since the measure $[\d B] e^{-S_{FP}[B]+\gamma^4\int d^4x h(B)}$ already ensures $\braket{h}=1$, there is no need for an additional critical parameter $\hat\gamma^4$ [more precisely, $\hat\gamma^4=0$], and we eventually arrive at
\begin{equation}\label{p5}
(\ast)= \int[\mathcal{D}B]  e^{-S_{FP}[B]+\gamma^4\int \d^4x h(B)}
\end{equation}
which is nothing else than the Gribov-Zwanziger action in its nonlocal formulation.

It should have not come as a surprise that imposing the somewhat more stringent Cucchieri-Mendes bound leads to exactly the Gribov-Zwanziger action after all. In the original derivation, the argument is that the dominant integration region is the boundary of $\Omega$. Naively speaking, imposing the Cucchieri-Mendes bound via \eqref{p1}, amounts to integrating over an outer shell of the ellipsoid $\Omega$. The dominant region will still be the outer boundary of this infinite-dimensional shell, being $\p\Omega$.

Let us here also point to \cite{Maas:2007uv,Cucchieri:2006tf,Sternbeck:2005vs} for a numerical verification that the relevant configurations are located close to $\p\Omega$, as it was found that the average value of the smallest nontrivial eigenvalue of the Faddeev-Popov operator approaches zero for growing lattice volume.

\section{Using the inverse ghost form factor as a ``nonperturbative gauge parameter'': construction of the corresponding action}
In this Section, we will take a closer look at the so-called Landau $B$-gauges as first introduced in \cite{Maas:2009se,Maas:2010wb}  and studied afterwards in \cite{Sternbeck:2012mf}, albeit mostly in a numerical lattice context. Obviously, the Landau gauge is assumed to begin with. The motivation lies in the Dyson-Schwinger quantum equations of motion (DSEs). These functional equations appear to have multiple solutions, and finding a unique solution depends on a choice of boundary conditions, see e.g.~\cite{Caianiello:1974pw,Bender:1988bp}. In the context of the DSEs, the (not always explicitly mentioned) boundary condition that is imposed, is making a choice for the $1PI$ ghost self energy at zero momentum, see e.g.~\cite[Sect V.C]{Aguilar:2011xe}. This observation was explicitly made and numerically tested in \cite{Watson:2010cn} for the ghost propagator in the Coulomb gauge.

More precisely, in the Landau gauge, using global color invariance, one may write
\begin{equation}\label{nnn1}
  \mathcal{G}^{ab}(p^2)=\frac{g(p^2)}{p^2}\delta^{ab}\,.
\end{equation}
One then imposes that the ghost form factor $g(p^2)$ obeys
\begin{equation}\label{nnn1bis}
g^{-1}(0)=B\,.
\end{equation}
A special case is provided by $B=0$, which corresponds to a ghost propagator that is more singular than $1/p^2$ in the deep infrared. This is commonly known as the ``scaling'' scenario, while $B>0$ corresponds to the ``decoupling/massive'' scenario, the latter being realized in current lattice simulations\footnote{At least one out of infinitely many decoupling solutions. Notice that we exclude $B<0$ by continuity reasons. For large $p^2$, $g^{-1}(p^2)>0$, and if $g^{-1}(0)<0$, we expect $g^{-1}(P^2)=0$, indicating a pole at some finite scale $P^2>0$, i.e.~a ghost tachyon. There is neither need nor any lattice evidence for such.}. This terminology was coined in \cite{Fischer:2008uz}. Although the question whether both scenarios could be realized on the lattice, is to our knowledge not yet answered with strict rigor, various reasons why finding a scaling solution on the lattice would be rather surprising can be found in the discussion of \cite{Cucchieri:2013nja}, see also \cite{LlanesEstrada:2012my} for an argumentation in the continuum. Interestingly, the latter paper also discusses the effective action related to both classes of solutions, giving evidence that the massive/decoupling solution corresponds to a more stable situation. The fact that multiple distinct solutions do exist, but that the theory itself singles one out on energetic grounds (smaller vacuum energy) sounds, evidently, physically appealing and logical. Arguments suggesting the possible existence of the scaling solution were provided in \cite{Sternbeck:2008mv}.

Our goal now is to show that it is possible to enforce the specific boundary condition $g^{-1}(0)=B$ directly at the level of the action, without making compromises w.r.t.~the renormalizability of the construction. This gives an explicit answer to the speculations of \cite{Maas:2009se,Maas:2010wb} if it is possible to implement at the functional level the boundary condition. Notice that being able to implement all constraints at the level of the Lagrangian/action allows for a clean discussion of the properties of the eventual solution. We will therefore combine the tools of \cite{Capri:2012wx,Dudal:2009xh}. Let us consider the Faddeev-Popov operator, $\mathcal{M}^{ab}$ and for the moment we treat $A_\mu$ as a classical external field. The ghost propagator in Fourier space reads
\begin{eqnarray}
\braket{ \bar{c}^a(p) c^b(-q)} &=& \int \d^d x \; \int \d^d y \braket{ \bar{c}^a(x)  c^b(y)} e^{-ipx}e^{iqy}\,,
\end{eqnarray}
with
\begin{eqnarray}
\braket{ \bar{c}^a(x) c^b(y)} &=& \frac{\int[{\cal D}c][{\cal D}\bar{c}] \;\bar{c}^a(x) c^b(y) e^{\int \bar{c}^c {\cal M}^{cd} c^d }}{\int[{\cal D}c][ {\cal D} \bar{c}]\;e^{\int \bar{c}^c {\cal M}^{cd} c^d}}\label{g2pf}\,.
\end{eqnarray}
For the ghost form factor, a compact expression can be derived if we look at the color trace of \eqref{g2pf}, that is
\begin{eqnarray}
\mathcal{G} (p,A) &=& \frac{1}{V(N^2-1)} \braket{\bar{c}^a(p) c^a(-q)}\delta(p+q)=\frac{1}{p^2} (1 + \sigma(p,A)) \label{gpara}\,,
\end{eqnarray}
with $V\equiv \delta(0)$ the (infinite) 4d spacetime volume. Using Wick's theorem, the transversality of the Landau gauge and some algebra, it was shown in \cite{Capri:2012wx} that $\sigma(p=0,A)$ can be compactly rewritten as
\begin{eqnarray}
\sigma(0, A) &=&   -\frac{g^2}{VD(N^2-1)} \int \frac{\d^D p}{(2\pi)^D} \int \frac{\d^D q}{(2\pi)^D}  A^{ab}_{\mu}(-p) \left({\cal M}^{-1}\right)^{bc}(p,q)A^{ca}_{\mu}(q)\equiv\frac{H(A)}{DV(N^2-1)}\,, \label{ff-hf}
\end{eqnarray}
where $A_\mu^{ab}=-f^{abc}A_\mu^c$ and $\mathcal{M}^{ab}(p,q)=  q^2\delta^{ab}\delta(p-q) - gA^{ab}_{\mu}(p-q)iq_{\mu}$.

Now, we can quantize the gauge field as well, and \eqref{ff-hf} can be translated into the following equality
\begin{eqnarray}
{\cal G} (k) &=& \langle {\cal G} (k,A)\rangle^{conn} = \frac{1}{k^2} (1 + \langle\sigma(k,A)\rangle^{conn}) =  \frac{1}{k^2} \frac{1}{(1 - \langle\sigma(k,A)\rangle^{1PI})}\label{np-gz}\,,
\end{eqnarray}
where ``$conn$'' stands for the connected set of diagrams and $1PI$  for the 1-particle irreducible ones. The boundary condition \eqref{nnn1bis} is then equivalent to requiring
\begin{eqnarray}\label{nnn2}
\braket{\sigma(0,A)}^{1PI}=1-B\equiv\hat B\,.
\end{eqnarray}
How to impose this at the level of the partition function? This can be accommodated for by a replacement of the Landau gauge Faddeev-Popov measure as follows:
\begin{eqnarray}\label{nnn3}
\int [\mathcal{D}\mu_{FP}]\equiv \int [\mathcal{D}A] \det\mathcal{M}\delta(\p A)e^{-S_{YM}}\to \int [\mathcal{D}\mu']&\equiv& \int [\mathcal{D}\mu_{FP}]\delta(\sigma(0)-\hat B) \\
&=& \int [\mathcal{D}A] \det\mathcal{M}\delta(\p A)\delta(H(A)-\hat B dV(N^2-1))e^{-S_{YM}}\nonumber\,.
\end{eqnarray}
In statistical mechanics terms, the latter measure corresponds to taking a microcanonical ensemble average. In the thermodynamic limit, $\sharp$~d.o.f.$~\to\infty$, $V\to\infty$; $\sharp$~d.o.f./$V$~=~constant; the measure can be reformulated with a steepest descent evaluation (aka.~saddle point approximation) to arrive at a Boltzmann ensemble\footnote{A clarifying illustrative exercise at tree level concerning the Boltzmann vs.~microcanonical equivalence can be found in \cite{Vandersickel:2012tz}.}. More precisely \cite{Gribov:1977wm,Dudal:2009xh},
\begin{eqnarray}\label{nnn4}
\int [\mathcal{D}\mu_{FP}]\delta(H(A)-\hat B dV(N^2-1))&=&\int [\mathcal{D}\mu_{FP}] \int_{-i\infty+\epsilon}^{i\infty+\epsilon}\frac{\d\beta}{2\pi i}e^{-\beta(H(A)-\hat B dV(N^2-1))}\nonumber\\
&=&\int_{-i\infty+\epsilon}^{i\infty+\epsilon}\frac{\d\beta}{2\pi i} e^{-E(\beta)}\,,
\end{eqnarray}
with
\begin{eqnarray}
E(\beta)=-\ln f(\beta)=-\ln \int [\mathcal{D}\mu_{FP}]e^{-\beta(H(A)-\hat B dV(N^2-1))}\,.
\end{eqnarray}
Power counting shows that $\beta$ has mass dimension $4$. To evaluate the $\beta$-integral of \eqref{nnn4}, we need to solve
\begin{equation}\label{nnn5}
\left.\frac{\p E(\beta)}{\p \beta}\right|_{\beta=\beta^*}=  \frac{f'(\beta^*)}{f(\beta^*)}=0
\end{equation}
to locate the critical point $\beta^*$ and make sure we can deform the contour to pass through $\beta^*$. This is allowed if the integration along the imaginary axis can be extended to a contour integration. From the representation \eqref{nnn4} it is actually clear that $E(\beta)$ corresponds to the vacuum energy up to the (infinite) volume factor $V$. On dimensional grounds, in $d$ dimensions, we will have $E(\beta)\propto \beta$, so if we close the contour on the right ($Re(\beta)>0$), the $\int e^{-E(\beta)}$ will vanish for $\beta\to\infty$ and we can safely deform the contour to pass through $\beta^*$, as long as $\beta^*\geq0$ and if there are no poles or cuts located in the complex $\beta$-plane with $Re(\beta)>0$. We will come back to the latter issue further on. A little algebra results in $\frac{\p^2 E}{\p \beta^2}=\braket{(H-\braket{H})^2}>0$, so we expect that $\frac{\p^2 E}{\p \beta^2}>0$ at the critical point. This will correspond to a global minimum of the vacuum energy\footnote{Notice that $E(0)=0$.}, and the steepest descent evaluation should thus be allowed.

Returning to the saddle point equation \eqref{nnn5}, we notice that it can be rephrased as
\begin{equation}\label{nnn6}
  \braket{H(A)}=\hat B dV(N^2-1)\,,
\end{equation}
where the expectation value is taken w.r.t.~the new measure
\begin{eqnarray}\label{nnn7}
\int [\mathcal{D}A] \det\mathcal{M}\delta(\p A)e^{-S_{YM}}e^{-\beta^*(H(A)-\hat B dV(N^2-1))}\,.
\end{eqnarray}
Now, since $E$ is composed of vacuum bubble contributions, any potential one-particle reducible diagram contributing to $E$ must be proportional to the square of the zero-momentum expectation value of the field propagating in the reducible line. This refers to nothing else than the vacuum expectation value of the fundamental field. Due to Lorentz and/or global color invariance, such condensates are however zero. Consequently, the relation \eqref{nnn6} actually reads
\begin{equation}\label{nnn8}
  \braket{H(A)}^{1PI}=\hat B dV(N^2-1)\,,
\end{equation}
which, upon using \eqref{ff-hf} again, leads to
\begin{equation}\label{nnn8}
  \braket{\sigma(0,A)}^{1PI}=\hat B\,,
\end{equation}
that is, we indeed have constructed a novel partition function that precisely gives the desired result, \eqref{nnn1bis}, $g^{-1}(0)=B$.

A word of caution is in order here. Whilst writing down \eqref{nnn3}, we made the a priori assumption that for each gauge configuration, there is at least one gauge equivalent configuration leading to the imposed value of $B$ while satisfying the Landau gauge condition\footnote{It would be no problem if some gauge fields are not included in the path integral, as long as it is a zero measure set. However, if relevant gauge fields were to be excluded, part of the (gauge invariant) physics would be missed.}. Unfortunately, there is no formal proof of this. To our knowledge, at most it is rigourously known that each gauge orbit has at least one representant inside the Gribov region $\Omega$, including its boundary \cite{Dell'Antonio:1991xt}. We will however stick to the assumption that every value of $B\geq0$ is allowed. This is in part motivated by the (less strong) assumption of the Dyson-Schwinger paper \cite{Fischer:2008uz} that, after taking expectation values, the ghost propagator's form factor can attain any $B>0$ as its zero momentum value. Further numerical studies concerning the range of allowed $B$'s and related discussion can be found in \cite{Maas:2011se,Maas:2013vd,Cucchieri:2013nja,Sternbeck:2012mf,Maas:2011ba}.

To bring the partition function in a more usual, i.e.~exponential, form, we can, completely analogously as done in the original Zwanziger derivation, introduce a set of auxiliary bosonic $\varphi_\mu^{ab}$,  $\overline\varphi_\mu^{ab}$, and fermionic fields, $\omega_\mu^{ab}$, $\overline\omega_\mu^{ab}$, to localize the nonlocal operator $H(A)$. We eventually find that the partition function is completely determined in terms of the action
\begin{eqnarray}\label{g2ff}
S &=& \frac{1}{4}\int \d^d x\; F^a_{\mu\nu} F^a_{\mu\nu}+\int \d^d x\,\left( b^a \p_\mu A_\mu^a +\overline c^a \p_\mu D_\mu^{ab} c^b \right)\nonumber\\&&+\int \d^d x \left( \overline{\varphi }_{\mu }^{ac} \partial _{\nu} D_\nu^{am}\varphi _{\mu
}^{mc} -\overline{\omega }_{\mu }^{ac} \partial _{\nu } D_\nu^{am} \omega _{\mu }^{mc}
  -g\left( \partial _{\nu }\overline{\omega }_{\mu}^{ac}\right) f^{abm}\left( D_{\nu }c\right) ^{b}\varphi _{\mu
}^{mc}\right)\nonumber\\
&&+\int\d^d x\left( -\sqrt{\beta^*}gf^{abc}A_{\mu }^{a}(\varphi _{\mu }^{bc}+\overline{\varphi }_{\mu }^{bc}) - d\hat B\left(N^{2}-1\right) \beta^* \right) \,.
\end{eqnarray}
Indeed, being Gaussian, both the $\varphi$- and $\omega$-integrals can be executed exactly, their respective determinants canceling due to their opposite Grassmann parity.

The mass scale $\beta^*$ is still to be fixed from the gap equation \eqref{nnn5}, which can, in the local formulation, also be expressed as
\begin{eqnarray}\label{localgap}
\braket{gf^{abc}A_{\mu }^{a}(\varphi _{\mu }^{bc}+\overline{\varphi }_{\mu }^{bc})}=2d\hat B(N^2-1)\sqrt{\beta^*}\,.
\end{eqnarray}
Obviously, the action we just constructed is a generalization of the Gribov-Zwanziger one, which is recovered in the limit $\hat B=1$, in which case the original no-pole condition of Gribov is implemented.

From the local formulation, to which we can apply the usual effective action formalism, it becomes evident that the vacuum energy, upon including the quantum corrections, will be a function of the type $E\left(\beta^\ast, g^2,\ln\frac{\sqrt{\beta^\ast}}{\mu^2}\right)$. Indeed, apart from the pure vacuum term, $\sqrt{\beta^\ast}$ only appears in the quadratic term that mixes the $A_\mu^a$-field with the $\varphi_\mu^{ac}$, $\overline\varphi_\mu^{ac}$ fields, thus it will only enter the (mixed) $(b^a,A_\mu^a, \varphi_\mu^{ac}, \overline\varphi_\mu^{ac})$ propagators. In the absence of any other scale, up to the global prefactor, the only way it can enter the vacuum energy is logarithmically. Taking as before $Re(\beta)>0$, we avoid the cut of the logarithm, and we can safely deform the contour for the saddle point evaluation.

By a simple rescaling $\beta^*\to\frac{\beta^*}{\hat B}$, the ``gauge parameter'' $\hat B$ will only enter via the $(b^a, A_\mu^a, \varphi_\mu^{ac}, \overline\varphi_\mu^{ac})$ propagators, a situation rather akin to a genuine gauge parameter. Though, the situation is very different, in particular can physical quantities, as $E$, explicitly depend on $\hat B$, see later in this paper.

\section{The inverse ghost form factor as a ``nonperturbative gauge parameter'': some quantum aspects}

\subsection{Renormalizability}
In order to discuss the renormalizability, we can follow the same argumentation as for the Gribov-Zwanziger action, as the only difference is the tree level pure vacuum term, which leads to a different gap equation.  We use the setup and notations of\footnote{Alternative routes to discuss the renormalization effects of a GZ-like theory can be found in \cite{Capri:2011wp,Capri:2013naa}. } \cite{Dudal:2010fq} which fits within the general algebraic formalism \cite{Piguet:1995er}. We set $\sqrt{\beta}=\gamma^2$ and first introduce a multi-index $i\equiv (\mu,a)$ which allows to rewrite the action \eqref{g2ff} as
\begin{eqnarray}\label{start}
S &=& S_{0} +  S_{\gamma} \,,
\end{eqnarray}
with
\begin{eqnarray}
S_{0}&=&S_\YM + S_\gf + \int \d^d x \left( \overline \varphi_i^a \p_\mu \left( D_\mu^{ab} \varphi^b_i \right) - \overline \omega_i^a \p_\mu \left( D_\mu^{ab} \omega_i^b \right) - g f^{abc} \p_\mu \overline \omega_i^a    D_\mu^{bd} c^d  \varphi_i^c \right) \nonumber \;, \\
S_{\gamma}&=& -\gamma ^{2}g\int\d^{d}x\left( f^{abc}A_{\mu }^{a}\varphi _{\mu }^{bc}  +f^{abc}A_{\mu}^{a}\overline{\varphi }_{\mu }^{bc} + \hat B\frac{d}{g}\left(N^{2}-1\right) \gamma^{2} \right) \,.
\end{eqnarray}
In order to find a sufficiently large set of Ward identities, it is useful to embed the previous action into another, more general one,
\begin{eqnarray}\label{brstinvariant}
\tilde\Sigma &=& S_0 + S_\s \,,
\end{eqnarray}
whereby
\begin{eqnarray}\label{previous}
S_\s &=& s\int \d^d x \left( -U_\mu^{ai} D_\mu^{ab} \varphi_i^b - V_\mu^{ai} D_{\mu}^{ab} \overline \omega_i^{b} - U_\mu^{ai} V_\mu^{ai}  + T_\mu^{a i} g f_{abc} D^{bd}_\mu c^d \overline \omega^c_i \right)\nonumber\\
&=& \int \d^d x \left( -M_\mu^{ai}  D_\mu^{ab} \varphi_i^b - gf^{abc} U_\mu^{ai}   D^{bd}_\mu c^d  \varphi_i^c   + U_\mu^{ai}  D_\mu^{ab} \omega_i^b - N_\mu^{ai}  D_\mu^{ab} \overline \omega_i^b - V_\mu^{ai}  D_\mu^{ab} \overline \varphi_i^b \right.\nonumber\\
&&\left.+ gf^{abc} V_\mu^{ai} D_\mu^{bd} c^d \overline \omega_i^c -\hat B M_\mu^{ai} V_\mu^{ai}+\hat B U_\mu^{ai} N_\mu^{ai}  + R_\mu^{ai} g f^{abc} D_\mu^{bd} c^d \overline \omega^c_i  + T_\mu^{ai} g f_{abc} D^{bd}_\mu c^d \overline \varphi^c_i\right) \,.
\end{eqnarray}
The Yang-Mills field are subject to their standard BRST variation,
\begin{align}\label{BRST1bis}
s A_\mu^a &= -D_{\mu}^{ab}c^b\,, & s c^a  &= \frac{g}{2}f^{abc}c^b c^c\,,   \nonumber \\
s \oc^a &= b^a\,, & s b^a &=0\,,
\end{align}
while the auxiliary fields are BRST doublets,
\begin{align}\label{BRST1bisb}
s \varphi_i^a  &= \omega_i^a\,, & s \omega_i^a  &= 0\,,   \nonumber \\
s \overline\omega_i^a &= \overline\varphi_i^a\,, & s \overline\varphi_i^a &=0\,.
\end{align}
We have also introduced 3 new BRST doublets of sources ($U_\mu^{ai}$, $M_\mu^{ai}$), ($V_\mu^{ai}$, $N_\mu^{ai}$) and ($T_\mu^{ai}$, $R_\mu^{ai}$) transforming according to
and
\begin{align}\label{BRST2}
sU_{\mu }^{ai} &= M_{\mu }^{ai}\,, & sM_{\mu }^{ai}&=0\,,  \nonumber \\
sV_{\mu }^{ai} &= N_{\mu }^{ai}\,, & sN_{\mu }^{ai}&=0\,,\nonumber \\
sT_{\mu }^{ai} &= R_{\mu }^{ai}\,, & sR_{\mu }^{ai}&=0\,.
\end{align}
For the special set of source values
\begin{eqnarray}\label{physlimit}
&& U_\mu^{ai} =  N_\mu^{ai} = T_\mu^{ai} = 0 \,, \nonumber\\
&& M_{\mu \nu }^{ab}= V_{\mu \nu}^{ab}=  R_{\mu \nu}^{ab} = \gamma^2 \delta ^{ab}\delta _{\mu \nu } \,,
\end{eqnarray}
we recover the action of interest, $S$, \eqref{start}. To consistently define the nonlinear BRST variations of $A_\mu^a$ and $c^a$, we also temporarily need to address
\begin{eqnarray}\label{volledig}
\Sigma &=& S_\sigma + \int d^4x \left(-K_\mu^a D_\mu^{ab}c^b+ \frac{g}{2}L^a f^{abc}c^b c^c\right)\,,
\end{eqnarray}
with
\begin{align}\label{BRST2}
sK_{\mu}^{a} &= 0\,, & sL^a&=0\,.
\end{align}
The ``full'' action $\Sigma$ obeys several powerful Ward identities. We refrain from writing them all down there, they can be found in the Appendices of \cite{Dudal:2010fq}; the only difference with the action discussed there is the presence of the additional parameter $\hat B$, which only affects the pure source (vacuum) term  $\propto~MV-UN$.  Clearly, a change of the latter will never have any consequence whatsoever on the UV structure of the theory, so the theory for any value of $\hat B$ will be renormalizable, given that the renormalizability has already been established in e.g.~\cite{Zwanziger:1992qr,Dudal:2010fq,Maggiore:1993wq} for $\hat B=1$.  In fact, there is an additional Ward identity directly related to $\hat B$, namely
\begin{eqnarray}
  \frac{\p \Sigma}{\p \hat B} &=& \int \d^4x\left(-M_\mu^{ai}V_\mu^{ai}  +U_\mu^{ai}N_\mu^{ai}\right)\,.
\end{eqnarray}
The latter is a pure source term, as such the identity remains valid at the quantum level for the $1PI$ effective action $\Gamma$. Consequently, as $\hat{B}$ does appear in the bare action but not in the counterterm, the renormalization of this pure source term, $\hat{B}(-MV + UN)$, will happen as
\begin{equation}
Z_{\hat{B}}\left(Z_{M}Z_{V}\hat{B}MV - Z_{U}Z_{N}\hat{B}UN\right) = \left(1-h \frac{a_{1}}{\hat{B}}\right)\hat{B}(-MV + UN)\,,
\end{equation}
where $h$ is the infinitesimal parameter associated to the quantum perturbation of the classical action \cite{Piguet:1995er}. As can be inferred from \cite{Dudal:2010fq}, $Z_{M}Z_{V}=Z_{U}Z_{N}=(1- h a_{1})=Z^{2}_{\gamma^{2}}$ and so we have
\begin{equation}
Z_{\hat{B}} = \left[1 - h a_{1} \left(\frac{1}{\hat{B}}-1\right)\right]\,,
\end{equation}
which finally gives
\begin{equation}
\label{zb}
Z_{\hat{B}} = \left(1 - h a_{1} \right)^{\left(\frac{1}{\hat{B}}-1\right)} = Z^{2\left(\frac{1}{\hat{B}}-1\right)}_{\gamma^{2}} \,.
\end{equation}
In the above, $a_{1}$ is an arbitrary parameter that comes from the counterterm and $Z_{\gamma^{2}}$, which is the multiplicative renormalization factor for $\gamma^2$, is given by $Z^{2}_{\gamma^{2}} = \left(1+\frac{3}{2}\frac{g^{2}N}{16\pi^{2}}\frac{1}{\varepsilon}\right)$ as can be learnt from \cite{Dudal:2008sp} (notations and conventions are following \cite{Dudal:2010fq}).

The only possible change in the other Ward identities of \cite{Dudal:2010fq} will at most occur in the pure source term, that is, in the potential harmless linear breaking terms (even zeroth order terms) of the classical/quantum Ward identities. For further usage, we will need the following Ward identities:
\begin{eqnarray}\label{wardje0}
  \frac{\delta \Sigma}{\delta b^a}&=& \p_\mu A_\mu^a \,,
\end{eqnarray}
\begin{eqnarray}\label{wardje1}
  \frac{\delta \Sigma}{\delta \omega_i^a}+\p_\mu \frac{\delta \Sigma}{\delta N_\mu^{ai}}-gf^{abc}\overline\omega_i^b\frac{\delta \Sigma}{\delta b^c}&=& -\hat B  \p_\mu U_\mu^{ai}+D_\mu^{ab}U_\mu^{ib}  \,,
\end{eqnarray}
and\footnote{This identity was actually not considered in \cite{Dudal:2010fq}.}
\begin{eqnarray}\label{wardje2}
\frac{\delta \Sigma}{\delta \overline\omega_i^a}+\p_\mu \frac{\delta \Sigma}{\delta U_\mu^{ia}}+gf^{abc}(V_\mu^{ci}+R_\mu^{ci})\frac{\delta \Sigma}{\delta K_\mu^b}&=& \hat B \p_\mu N_\mu^{ai}-D_\mu^{ab}N_\mu^{bi}\,,
\end{eqnarray}
next to the global invariance (thanks to the multi-index)
\begin{eqnarray}
U_{ij}\Sigma&=&0\\
U_{ij}&=&\int \d^dx\Bigl( \varphi_{i}^{a}\frac{\delta }{\delta \varphi _{j}^{a}}-\overline{\varphi}_{j}^{a}\frac{\delta }{\delta \overline{\varphi}_{i}^{a}}+\omega _{i}^{a}\frac{\delta }{\delta \omega _{j}^{a}}-\overline{\omega }_{j}^{a}\frac{\delta }{\delta \overline{\omega }_{i}^{a}} -  M^{aj}_{\mu} \frac{\delta}{\delta M^{ai}_{\mu}} -U^{aj}_{\mu}\frac{\delta}{\delta U^{ai}_{\mu}} + N^{ai}_{\mu}\frac{\delta}{\delta N^{aj}_{\mu}}\nonumber\\&&  +V^{ai}_{\mu}\frac{\delta}{\delta V^{aj}_{\mu}}    +  R^{aj}_{\mu}\frac{\delta}{\delta R^{ai}_{\mu}} + T^{aj}_{\mu}\frac{\delta}{\delta T^{ai}_{\mu}} \Bigr)  \,, \nonumber
\end{eqnarray}
by means of which we can introduce the ``$i$-charge'' $Q=U_{ii}$. Last but not least, we will also need the Slavnov-Taylor identity:
\begin{eqnarray}\label{ST}
  \int d^4x\left(\frac{\delta \Sigma}{\delta K_\mu^a}\frac{\delta \Sigma}{\delta A_\mu^a}+\frac{\delta \Sigma}{\delta L^a}\frac{\delta \Sigma}{\delta c^a}+b^a\frac{\delta \Sigma}{\delta \overline c^a}+\overline\varphi_i^a\frac{\delta \Sigma}{\delta \overline\omega_i^a}+\omega_i^a\frac{\delta \Sigma}{\delta \varphi_i^a}+M_\mu^{ai}\frac{\delta \Sigma}{\delta U_\mu^{ai}}+N_\mu^{ai}\frac{\delta \Sigma}{\delta V_\mu^{ai}}+R_\mu^{ai}\frac{\delta \Sigma}{\delta T_\mu^{ai}}\right)  &=& 0\,.
\end{eqnarray}

\subsection{The anomalous dimension $\gamma_{\hat{B}}$}
It is interesting to take a closer look at the anomalous dimension of $\hat{B}$, which we define via
\begin{equation}
\gamma_{\hat{B}}= - \frac{\p \ln Z_{\hat{B}}}{\p \ln \bar{\mu}}
\end{equation}
and from \eqref{zb} we learn
\begin{equation}
\ln Z_{\hat{B}} = \left(\frac{1}{\hat{B}}-1\right)\frac{3}{2}\frac{\alpha}{\varepsilon}\;, \qquad \text{with} \qquad \alpha = \frac{g^{2}N}{(4\pi)^{2}}\,.
\end{equation}
Therefore we can write
\begin{equation}
\gamma_{\hat{B}}= - \frac{\p \ln Z_{\hat{B}}}{\p \alpha} \frac{\p \alpha}{\p \ln \bar{\mu}}\,.
\end{equation}
Setting $g_{0}^{2} = \bar{\mu}^{\varepsilon}Z^{2}_{g}g^{2}$ we have
\begin{equation}
\frac{\p \alpha}{\p \ln \bar{\mu}} = \left(-\varepsilon g^{2}+2\beta(g)\right)\frac{N}{(4\pi)^{2}}\,,
\end{equation}
so that
\begin{equation}
\gamma_{\hat{B}}= - \left(-\varepsilon g^{2}+2\beta(g)\right)\frac{N}{(4\pi)^{2}} \left(\frac{1}{\hat{B}}-1\right)\frac{3}{2}\frac{1}{\varepsilon}\,,
\end{equation}
leading to
\begin{equation}
\label{gammab}
\gamma_{\hat{B}}= \left(\frac{1}{\hat{B}}-1\right)\frac{3}{2}\frac{g^{2}N}{(4\pi)^{2}}\,.
\end{equation}
According to \eqref{gammab} and since $\hat B \geq 1$ the anomalous dimension of $\hat{B}$ is thus positive and it has a fixed point at $\hat{B}=1$, reached in the UV regime. The fact that $\hat B=1$ is a fixed point to all orders\footnote{This can be easily inferred from its renormalization factor, \eqref{zb}.} is a cross-check that the original Gribov-Zwanziger action is a stable special case of the general action that gives a prescribed zero momentum ghost form factor. Since $\hat B=1$ is reached in the ultraviolet regime, it would appear that, starting from a choice of $\hat B$ that does not correspond to the original Gribov-Zwanziger case (scaling solution), one will likely not end up with $\hat B=1$ at low momentum.

\subsection{Explicit proof of the boundary condition being realized}
In this subsection, we provide a proof that the ghost propagator indeed obeys the boundary condition \eqref{nnn1bis} when quantum corrections are taken into account. The argumentation will solely rest on the underlying Ward identities defining the quantum effective action. In the case of $\hat B=1$, a proof can already be found in \cite{Zwanziger:1992qr}, but the here presented more general new proof is shorter and somewhat less technical and applies to general $\hat B$. The gain is coming from an explicit use of the quantum Slavnov-Taylor identity.

We depart from the partition function associated with the action \eqref{g2ff}. By adding sources coupled to the Faddeev-Popov ghosts and Gribov-Zwanziger ghosts, integrating them out exactly as they appear at most quadratically, one finds the general identification (see e.g.~Section~III of \cite{Dudal:2010fq}, albeit the proportionality factor mentioned there is now corrected here.).
\begin{eqnarray}\label{identi}
  \braket{f(A) (\mathcal{M}^{-1})^{ab}(x,y)} &=& \braket{f(A) c^a(x)\overline c^b(y)}=-\frac{1}{4(N^2-1)}\braket{f(A) \omega_i^a(x) \overline \omega_i^b(y)}\,,
\end{eqnarray}
where $f(A)$ is a generic functional of the gauge fields; the $4(N^2-1)$ corresponds to $\delta_{ii}$, where $i$ runs over $d(N^2-1)$ values, with $d$ the space-time dimensionality. Rather than to the ghost form factor at zero momentum  itself, we can thus equally well look at the equivalent expression
\begin{eqnarray}\label{identi2}
g(0)= -\frac{1}{4(N^2-1)^2} \lim_{p^2\to0} p^2 \int \frac{\d^4x}{(2\pi)^4} e^{ip(x-y)} \Braket{\omega_i^a(x) \overline\omega_i^a(y)}\,.
\end{eqnarray}
For now, let us focus on the Green function $\Braket{\omega_i^a(x) \overline\omega_i^a(y)}$, or even more specifically, to its $1PI$ counterpart (inverse propagator). The latter can be inferred from
\begin{equation}\label{identi3}
  \left.\frac{\delta^2 \Gamma}{\delta \omega_i^a(x) \delta \overline\omega_j^c(y)}\right|_0\,,
\end{equation}
where after the derivation, all (external) fields are put to zero, while also the sources should be attributed their physical value \eqref{physlimit}, both manipulations denoted with $\left.\right|_0$. $\Gamma$ is the quantum effective action, which is subject to the same constrains (Ward identities) as its classical version $\Sigma$. In particular, acting with $\frac{\delta}{\delta\overline\omega_j^c}$ on
\begin{eqnarray}\label{identi4}
  \frac{\delta \Gamma}{\delta \omega_i^a(x)}+\p_\mu^x \frac{\delta \Gamma}{\delta N_\mu^{ai}(x)}-gf^{abc}\overline\omega_i^b(x)\frac{\delta \Gamma}{\delta b^c(x)}&=& -\hat B  \p_\mu^x U_\mu^{ai}(x)+D_\mu^{ab,x}U_\mu^{ib}(x)
\end{eqnarray}
gives
\begin{equation}\label{identi5}
  \left.\frac{\delta^2 \Gamma}{\delta \omega_i^a(x) \delta \overline\omega_j^c(y)}\right|_0=\left.-\p_\mu^x \frac{\delta^2 \Gamma}{\delta N_\mu^{ai}(x)\delta\overline\omega_j^c(y)}\right|_0\,,
\end{equation}
since we also have $\frac{\delta \Gamma}{\delta b^a}=\partial_\mu A_\mu^a$. The previous expression \eqref{identi5} can be further simplified, using the quantum version of the EOM \eqref{wardje2}:
\begin{equation}\label{identi6}
  \left.\frac{\delta^2 \Gamma}{\delta \omega_i^a(x) \delta \overline\omega_j^c(y)}\right|_0=-\left.\p_\mu^x \p_\nu^y\frac{\delta^2 \Gamma}{\delta N_\mu^{ai}(x)\delta U_\nu^{cj}(y)}\right|_0+(1-\hat B)\p_\mu^x \p_\mu^y\delta_{ij}\delta_{ac}\delta(x-y)\,,
\end{equation}
while we keep the relations \eqref{physlimit} in mind, together with global color invariance. Notice that, since there are no $(\omega,\overline c)$-vertices, the contribution of the term $\propto \frac{\delta^2\Gamma }{\delta K \delta N}$ will vanish, as it corresponds to insertions of operators $\sim c\overline\omega$ and no diagrams can be formed of this type.

Given $\tilde\Sigma$ as in \eqref{brstinvariant}, \eqref{identi6} actually corresponds to
\begin{equation}\label{identi6b}
  \left.\frac{\delta^2 \Gamma}{\delta \omega_i^a(x) \delta \overline\omega_j^c(y)}\right|_0=-\p_\mu^x \p_\nu^y \braket{D_\mu^{ak}\overline\omega_i^k(x) D_\nu^{cm}\omega_j^m(y)}_{1PI} +\p_\mu^x \p_\mu^y\delta_{ij}\delta_{ac}\delta(x-y)\,.
\end{equation}
We also need to reformulate the gap equation \eqref{localgap} a bit. Reconsider the relevant (physical) action \eqref{start}. First we notice that the term $\mathcal{T}\equiv - g f^{abc} \p_\mu \overline \omega_i^a    D_\mu^{bd} c^d  \varphi_i^c$ actually plays no role when computing quantities with Grassmann charge zero; indeed, there are no compensating $(\omega,\overline c)$-vertices that would allow for the vertices of $\mathcal{T}$ to generate any diagram. So, for practical purposes we can actually set $\mathcal{T}=0$. Doing so, the action \eqref{start} enjoys the $\varphi\leftrightarrow\overline\varphi$ symmetry; the extra $\overline\varphi \varphi\p A$-term which comes about after interchanging the role of $\varphi$ and $\overline\varphi$ can be easily overcome by a shift in the multiplier $b$-field, or equivalently said, it vanishes on-shell because of the Landau gauge $\p A=0$. This observation can be used to reformulate the boundary condition \eqref{localgap}
\begin{eqnarray}\label{localgap2}
\braket{gf^{abc}A_{\mu }^{a}\varphi _{\mu }^{bc}}=\braket{gf^{abc}A_{\mu }^{a}\overline{\varphi }_{\mu }^{bc}}=\frac{1}{2}\braket{gf^{abc}A_{\mu }^{a}(\varphi _{\mu }^{bc}+\overline{\varphi }_{\mu }^{bc})}=d\hat B(N^2-1)\gamma^2\,.
\end{eqnarray}
Next, we notice that
\begin{eqnarray}\label{localgap3}
d\hat B(N^2-1)\gamma^2=\braket{gf^{abc}A_{\mu }^{a}\overline{\varphi }_{\mu }^{bc}}= \braket{D_\mu^{bc}\overline\varphi_\mu^{bc}}\,.
\end{eqnarray}

Nextly, we consider the quantum Slavnov-Taylor identity,
\begin{eqnarray}\label{ST2}
  \int \d^4x\left(\frac{\delta \Gamma}{\delta K_\mu^a}\frac{\delta \Gamma}{\delta A_\mu^a}+\frac{\delta \Gamma}{\delta L^a}\frac{\delta \Gamma}{\delta c^a}+b^a\frac{\delta \Gamma}{\delta \overline c^a}+\overline\varphi_i^a\frac{\delta \Gamma}{\delta \overline\omega_i^a}+\omega_i^a\frac{\delta \Gamma}{\delta \varphi_i^a}+M_\mu^{ai}\frac{\delta \Gamma}{\delta U_\mu^{ai}}+N_\mu^{ai}\frac{\delta \Gamma}{\delta V_\mu^{ai}}+R_\mu^{ai}\frac{\delta \Gamma}{\delta T_\mu^{ai}}\right)  &=& 0\nonumber\\
\end{eqnarray}
and we act on it with $\frac{\delta}{\delta N_\mu^{ai}}$, then we set all external fields and sources (except for $U$, $M$, $V$, $N$, $R$, $T$) to zero, thereby finding the identity
\begin{eqnarray}\label{ST3}
  &&\int \d^4 y \left(M_\nu^{cj}(y)\frac{\delta^2 \Gamma}{\delta N_\mu^{ai}(x)\delta U_\nu^{cj}(y)}\right)+\frac{\delta \Gamma}{\delta V_\mu^{ai}(x)}\nonumber\\&&- \int \d^4 y\left(N_\nu^{cj}(y)\frac{\delta^2\Gamma}{\delta N_\mu^{ai}(x)\delta V_\nu^{cj}(y)}+R_\nu^{cj}(y)\frac{\delta^2 \Gamma}{\delta N_\mu^{ai}(x)T_\nu^{cj}(y)}\right)= 0\,.
\end{eqnarray}
Subsequently, we assign to the remaining sources their physical values. The contributions from the second line vanish, since $N=0$ and because we can also drop the term $\sim\frac{\delta^2 \Gamma}{\delta N \delta T}$ since it corresponds to a correlation function of the type $\braket{\ldots \overline\omega c}^{1PI}$, which is yet again zero since no diagrams can be formed because of the absence of $(\omega,\overline c)$ vertices. We also have
\begin{eqnarray}\label{ST4}
  -\frac{\delta \Gamma}{\delta V_\mu^{ai}} &=& \braket{D_\mu^{ab}\overline \varphi_i^b}^{1PI}+BM_\mu^{ai}\,.
\end{eqnarray}
Substituting \eqref{physlimit} and \eqref{ST4} into \eqref{ST3}, we obtain, taking into account the expression for $\Sigma$ and the gap equation \eqref{localgap3}, while also keeping in mind the underlying global color and $Q_i$-charge invariance,
\begin{equation}\label{ST5}
\int \d^d y \braket{(D_\mu^{ab}\overline\omega_\mu^{ab})_x (D_\nu^{cd}\omega_\nu^{cd})_y}^{1PI}=d\hat B(N^2-1)\,.
\end{equation}
The $1PI$-ness is of course coming from the fact we worked with the $1PI$ generating functional $\Gamma$. The limit $d\to4$ is implicitly understood.

In momentum space, \eqref{ST5} can be reformulated as a condition on the ($1PI$) zero momentum correlation function:
\begin{equation}\label{localgap6}
\lim_{p^2\to0}\braket{(D_\mu^{ab}\overline\omega_\mu^{ab}) (D_\nu^{cd}\omega_\nu^{cd})_p}^{1PI}=d\hat B(N^2-1)\,.
\end{equation}
Specifically, parameterizing the $1PI$ correlation function as
\begin{equation}\label{localgap11}
\braket{(D_\mu^{ab}\overline\omega_i^{b}) (D_\nu^{cd}\omega_j^{d})}^{1PI}_p=\delta_{ij}\delta_{ac}\left(\mathcal{F}_1(p^2)\delta_{\mu\nu}+\mathcal{F}_2(p^2)p_\mu p_\nu\right)\,,
\end{equation}
we get from \eqref{localgap6}  that
\begin{equation}\label{localgap12}
  \mathcal{F}_1(0)=\hat B\,,
\end{equation}
at least when assuming that $\mathcal{F}_2(p^2)$ does not develop a dynamical pole at zero momentum.  That the latter would be rather unlikely has been motivated for by a Dyson-Schwinger analysis in \cite{Aguilar:2009pp}.

Finally, inserting \eqref{identi6b} into \eqref{identi2}, we get
\begin{eqnarray}\label{identi2}
g(0)= \frac{1}{1-\mathcal{F}_1(0)}=\frac{1}{1-\hat B}=\frac{1}{B}\,,
\end{eqnarray}
that is, the envisaged boundary condition.

\subsection{Explicit one-loop analysis}
It is instructive to check if the the boundary condition for the ghost propagator is indeed satisfied at the first order in the loop expansion. For this, we will need the vacuum energy and its behaviour with respect to $\gamma^{2}$ and $\hat{B}$. With this aim in mind, we only need the quadratic terms of \eqref{start} in order to obtain the vacuum energy and the ghost propagator. As some renormalization scheme will be used, the distinction between the bare and renormalized quantities will be explicitly written down. As usual the bare quantities will be identified by a subscript zero: as $\phi_{0}$ for denoting fields and $\lambda_{0}$ for denoting parameters. There are no specific labels for renormalized fields and parameters. The partition function accounting only for the quadratic terms of the proposed action can be recast in
\begin{equation}
Z_{quad} = \int [{\cal D}A[[{\cal D}c][{\cal D}\bar{c}] \exp\left[
-\int \frac{\d^{d}p}{(2\pi)^{d}}\, \left\{
\frac{1}{2}A^{a}_{\mu}(p) Q^{ab}_{\mu\nu}A^{b}_{\nu}(-p)
- \bar{c}^{a}(p) p^{2} c^{b}(-p) - \gamma^{4}\hat{B}d(N^{2}-1)
\right\}
\right]\;,
\end{equation}
where
\begin{equation}
Q^{ab}_{\mu\nu} = \frac{1}{2}p^{2}\left[ \left(1+\frac{2\gamma^{4}g^{2}N}{p^{4}}\right)\delta_{\mu\nu}-\left(1-\frac{1}{\alpha}\right)\frac{p_{\mu}p_{\nu}}{p^{2}}\right]\delta^{ab}\;.
\end{equation}
When the limit $\alpha \to 0$ is taken at the end of a computation, the Landau gauge is recovered.

\begin{figure}[t]
  \centering
  \scalebox{0.5}{\includegraphics{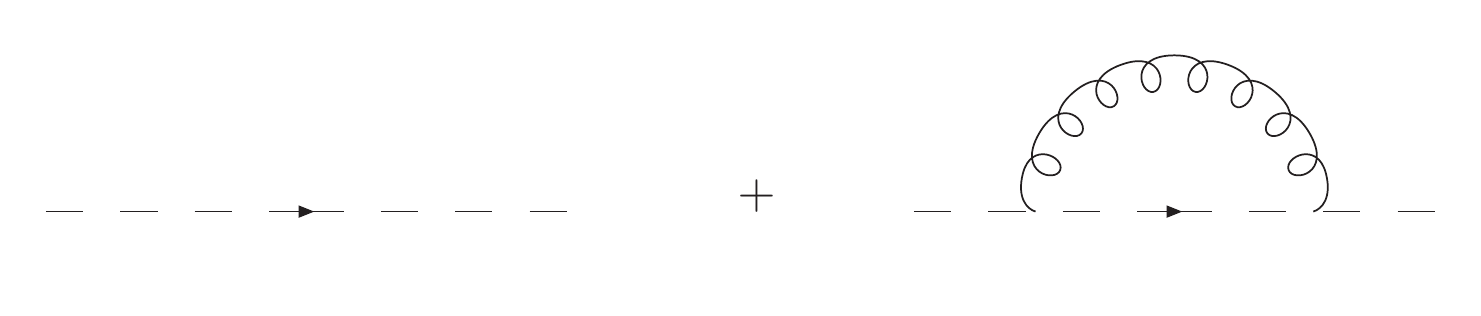}}
  \caption{Feynman diagram of ghost propagator at one-loop order.}\label{figure2}
\end{figure}

Therefore, as the (bare) vacuum energy is defined by
\begin{equation}
e^{-VE} = Z\;,
\end{equation}
in the general case, we are able to find
\begin{equation}
e^{-VE} = \exp\left\{ \gamma_{0}^{4}V\hat{B}_{0}d(N^{2}-1) - \frac{1}{2} \text{Tr} \ln Q^{ab}_{\mu\nu} + \text{Tr} \ln p^{2}\right\}\,,
\end{equation}
which gives,
\begin{equation}
  E= - \gamma_{0}^{4}\hat{B}_{0}d(N^{2}-1) +\frac{1}{2}(N^{2}-1)(d-1)\int \frac{\d^{d}p}{(2\pi)^{d}}\,\ln \left(\frac{p^{4}+2\gamma^{4}Ng^{2}}{p^{2}}\right)-\int \frac{\d^{d}p}{(2\pi)^{d}}\,\ln p^{2}\,.
\end{equation}
For the ghost propagator, see Figure \ref{figure2}, we need just to calculate the ghost form factor $\langle\sigma(0)\rangle$, as can be seen form \eqref{gpara}, \eqref{ff-hf} and \eqref{np-gz}, keeping in mind that
\begin{equation}
\langle A^{a}_{\mu}(p)A^{b}_{\nu}(-p)\rangle = \frac{p^{2}}{p^{4}+2\gamma^{4}g^{2}N}P_{\mu\nu}\delta^{ab}\;.
\end{equation}
Then we have
\begin{equation}
\langle\sigma(0)\rangle = \frac{g_{0}^{2}N}{V(N^{2}-1)}\frac{1}{p^{2}} \int \frac{\d^{d}k}{(2\pi)^{d}}\, \frac{(p-k)_{\mu}p_{\nu}}{(p-k)^{2}}\frac{k^{2}}{k^{4}-2g^{2}N\gamma^{4}}\left(\delta_{\mu\nu}-\frac{k_{\mu}k_{\nu}}{k^{2}}\right)\,,
\end{equation}
which can be reduced to
\begin{equation}
\langle\sigma(0)\rangle = \frac{g_{0}^{2}N}{V(N^{2}-1)}\frac{(d-1)}{d} \int \frac{\d^{d}k}{(2\pi)^{d}}\, \frac{1}{k^{4}-2g^{2}N\gamma^{4}}\,.
\end{equation}
Making use of $\MSbar$ renormalization scheme after dimensional regularization, with $d = 4-\varepsilon$ ($\varepsilon \to 0$) and with the result \eqref{zb}, we are led to following renormalized quantities,
\begin{equation}\label{vac1}
  E=-\frac{2(N^2-1)}{g^2N}\hat B\lambda^4+\frac{3(N^2-1)}{64\pi^2}\lambda^4\left(\frac{8}{3}-2\ln\frac{\lambda^2}{\omu^2}\right)\,,
\end{equation}
for the vacuum energy, and
\begin{equation}
\label{gffct}
\langle \sigma (0) \rangle = \frac{3}{4}\frac{g^{2}N}{(N^{2}-1)}\frac{1}{(4\pi)^{2}}\left(\frac{5}{6}-\ln \frac{\lambda^{2}}{\bar{\mu}^{2}}\right)
\end{equation}
for the ghost form factor, upon setting $\lambda^4=2g^2N\beta$.

In order to evaluate \eqref{gffct} we need to fix the value of $\lambda^{2}$ ($\sim \sqrt{\beta}$), which is done by solving the gap equation,
\begin{equation}\label{vac2}
0=\left.\frac{\p E}{\p \lambda^2}\right|_{\lambda^2=\lambda^{*2}}= -\frac{4(N^2-1)}{g^2N}\hat B \lambda^{*2}+\frac{3(N^2-1)}{32\pi^2}\lambda^{*2}\left(\frac{8}{3}-2\ln\frac{\lambda^{*2}}{\omu^2}\right)-\frac{3(N^2-1)}{32\pi^2}\lambda^{*2}=0\,,
\end{equation}
leading to the following vacuum energy and ghost form factor respectively,
\begin{equation}\label{vac3}
  E=\frac{3(N^2-1)}{64\pi^2}\lambda^{*4}
\end{equation}
and
\begin{equation}\label{gffct2}
\langle \sigma (0) \rangle = \hat{B}\,.
\end{equation}
Therefore, the boundary condition over the ghost propagator at first loop order was explicitly obtained from the proposed action \eqref{start}.

Interestingly enough, the second derivative at the critical value is negative,
\begin{equation}\label{vac3b}
\left.\frac{\p^2 E}{(\p \lambda^2)^2}\right|_{\lambda^2=\lambda^{*2}}=-\frac{3(N^2-1)}{16\pi^2}\,,
\end{equation}
meaning that we are dealing with a local maximum rather than minimum, putting in a different perspective the comments made below \eqref{nnn5}. Notice that the general argumentation made there and the explicit one loop verification are not necessarily at odds with each other. An important subtlety not taken into account below \eqref{nnn5} is, besides that in practice we work in a formal series expansion, one must renormalize, which means infinities are subtracted. It is immediate to realize a positive infinity can become negative after the renormalization procedure. The derivation of the action and gap equation must be understood at the formal and bare level, in which case there is no problem with negative variance $\braket{(H-\braket{H})^2}$.

The gap equation \eqref{vac2} can be explicitly solved for by summing the potentially large logarithms. This is achieved by setting $\omu^2=\lambda^{*2}$ (corresponding to use the renormalization group invariance of the effective action), giving
\begin{equation}\label{vac4}
  \frac{g^2N}{16\pi^2}=\frac{8}{5}\hat B=\frac{1}{\beta_0\ln\frac{\lambda^2}{\lms^2}}\,,
\end{equation}
where the one-loop coefficient of the $\beta$-function reads $\beta_0=\frac{11}{3}\frac{N}{16\pi^2}$. Henceforth,
\begin{equation}\label{vac5}
  \lambda^{*2}=\lms^2 e^{\frac{15}{88\hat B}}\,,\qquad  E=\frac{3(N^2-1)}{64\pi^2}\lms^4 e^{\frac{15}{44\hat B}}\,.
\end{equation}
Having found the expression for the vacuum energy in terms of the parameter $\hat B$, one might wonder if it would be possible to find a specific value for $\hat B$ by using it as a variational parameter. In the CJT formalism \cite{Cornwall:1974vz}, the two-point functions are to be determined by minimizing the effective action w.r.t.~those\footnote{A closely related exercise was carried out \cite{LlanesEstrada:2012my} with the $2PI$ effective action formalism to discriminate between different solutions of the Landau gauge Dyson-Schwinger equations.
}. However, this frequently amends to a tremendously difficult task, in which case it can be fortuitous to model the propagators in terms of a (few) parameter(s), leading to algebraic equations for these parameters from the minimization of the effective action. From this perspective, we can indeed employ $\hat B$ as a variational parameter associated to a specific ghost propagator and discuss whether some $\hat B$ minimizes the corresponding vacuum functional.

In a one-loop approximation, given the result \eqref{vac5}, it is immediately clear that $\frac{\p E}{\p \hat B}=0$ has no sensible solutions in terms of $\hat B$. This is not a coincidence, as we can provide an exact argument valid at any order. Writing the vacuum functional as
\begin{equation}\label{vac7}
  E(\lambda^2)=\tilde E(\lambda^2)-\frac{2(N^2-1)}{g^2N}\hat B\lambda^4\,,
\end{equation}
we need to solve, for each value of $\hat B$,
\begin{equation}\label{vac8}
0=\left.\frac{\p E}{\p \lambda^2}\right|_{\lambda^2=\lambda^{*2}}= -\frac{4(N^2-1)}{g^2N}\hat B \lambda^{*2}+\left.\frac{\p \tilde E_{vac}}{\p \lambda^2}\right|_{\lambda^2=\lambda^{*2}}\,.
\end{equation}
The vacuum energy, $E(\lambda^{*2})$, must then depend minimally on $\hat B$. The condition can be written as
\begin{equation}\label{vac9}
0=\left.\frac{\p E}{\p \hat B}\right|_{\lambda^2=\lambda^{*2}}=\left.\frac{\p \tilde E}{\p \lambda^{2}}\right|_{\lambda^2=\lambda^{*2}}\frac{\p \lambda^{*2}}{\p \hat{B}}-\frac{4(N^2-1)}{g^2N}\hat B\lambda^{*2}\frac{\p \lambda^{*2}}{\p \hat{B}}-\frac{2(N^2-1)}{g^2N}\lambda^{*4}\,.
\end{equation}
Upon using the supplementary condition \eqref{vac8}, the expression \eqref{vac9} reduces to
\begin{equation}\label{vac10}
-\frac{2(N^2-1)}{g^2N}\lambda^{*4}=0\,,
\end{equation}
requiring that $\lambda^{*2}=0$, which is not allowed, otherwise the gauge field region of integration will not be restricted to the first Gribov region and, thus, the ambiguity problem in gauge fixing will be present.  This means that using our formalism, there is no optimal value for $\hat B$ to be found that would minimize the vacuum energy, at least when restricting the analysis to a specific order in perturbation theory.

\section{Discussion}
In the first part of this paper, we have tried to take into account at the level of the path integration the recently derived Cucchieri-Mendes bounds on the ghost propagator in the presence of an external gauge field that belongs to the Gribov region $\Omega$. Our result indicates that this bound might not leave its footprints in the continuum restriction of the gauge field path integral, since the integral gets its main contribution from gauge fields close to the boundary where the (stricter) bound of Cucchieri-Mendes is saturated. Our result is based on an infinite volume continuum formulation, so we cannot say much about what happens exactly when a genuine discrete setting (as on a lattice, the setting of \cite{Cucchieri:2013nja}) would be used, e.g.~how ``fast'' the boundary of the Gribov region is approached or how this would depend on a variable lattice size/volume.

In the second part of the paper, we have provided a positive answer to the query put forward by Maas in \cite{Maas:2009se,Maas:2010wb} by constructing an action principle that allows to give a prescribed value to the zero momentum ghost form factor and this without jeopardizing the renormalizability. We have also given a formal proof, based on the Ward identities of the theory, that the boundary condition is met.
One issue we did not touch upon is that we did not show that our implementation of the restriction corresponds to a stable theory. Likely, following \cite{Dudal:2007cw,Dudal:2008sp,Dudal:2011gd,Gracey:2010cg}, the vacuum will be unstable against the condensation of $d=2$ operators which will drastically effect the properties of e.g.~the propagators, given the underlying vacuum will get nonperturbatively changed. If this happens, the Ward identities of the new theory will be different, and the proof that the boundary condition is realized, will no longer apply. Said otherwise, the ghost form factor at zero momentum, $g(0)$, will attain a value that will be not predetermined, but needs to be computed in one or another approximation. This scenario would correspond to a quantum modification of the fact that the path integral is geometrically dominated by configurations near to the boundary $\p\Omega$ of the Gribov region.

An important ingredient in the refinement of Gribov-Zwanziger like theories is the condensation of the $d=2$ operator $\overline\varphi_\mu^{ab}\varphi_\mu^{ab}-\overline\omega_\mu^{ab}\omega_\mu^{ab}=-s(\overline\omega_\mu^{ab}\varphi_\mu^{ab})$. This operator can be coupled to the action $\Sigma$, \eqref{volledig}, via the addition of $-\int \d^dx J(\overline\varphi_\mu^{ab}\varphi_\mu^{ab}-\overline\omega_\mu^{ab}\omega_\mu^{ab})$ with $J$ a local BRST invariant source, $sJ=0$. The used Ward identities will thus at most get modified by linear breakings in $J$, so they remain useful at the quantum level. The main obstacle, however, is that we loose the connection \eqref{identi} between the Faddeev-Popov ghost propagator and the auxiliary Gribov-Zwanziger ghost propagator, since the latter acquire a dynamical mass due to $J$ (more precisely, due to the ensuing condensation, see \cite{Dudal:2007cw,Dudal:2008sp,Dudal:2011gd}). So, even if we would be able to prove something about the $\braket{\overline\omega\omega}$ propagator at zero momentum, we do not longer have a connection to the propagator of interest, $\braket{\overline c c}$. Also in functional formalisms, a similar caveat might interfere. One can write down the quantum (Dyson-Schwinger) equations of motion based on the action \eqref{g2ff}. Since the Ward identities should in principle be obeyed by the nonperturbative $n$-point functions\footnote{Making abstract of the fact that one should be careful ---from the perspective of respecting these identities-- about making assumptions on truncating the a priori infinite tower of Dyson-Schwinger equations or modelling in higher $n$-point functions.}, the boundary condition of the ghost propagator should remain valid. Though, if composite operators, as $\overline\varphi_\mu^{ab}\varphi_\mu^{ab}-\overline\omega_\mu^{ab}\omega_\mu^{ab}$ can condense, their dynamics can be included in the functional formalism as well, e.g.~by applying the CJT method \cite{Cornwall:1974vz}. Similarly, the Ward identities of the CJT action will be adapted, changing the conclusion about the ghost boundary condition.

Lastly, an important feature of our formalism, completely analogous to the standard Gribov-Zwanziger formalism, is that it comes with a soft breaking of the BRST. Indeed, taking for instance the $s$-variation of the action \eqref{g2ff}, we get
\begin{eqnarray}\label{g2ffb}
sS &=& \int d^4x\left(\sqrt{\beta^*} gf^{abc}D_\mu^{am}c^m(\varphi_\mu^{bc}+\overline\varphi_\mu^{bc})-\sqrt{\beta^*}gf^{abc}A_\mu^a\omega_\mu^{bc}\right)
\end{eqnarray}
which is clearly nonzero if $\beta^*\neq0$. In retrospect, this also indicates we cannot give $\hat B$ the same meaning as a genuine gauge parameter. Clearly, physical quantities will depend on it. A first explicit example is the vacuum energy that varies with $\hat B$. If we define physical operators as the quantum extensions of the classical gauge invariant operators (see e.g.~\cite{Dudal:2009zh}), their expectation values will in general depend on $\hat B$. First of all, $\hat B$ is not coupled to a BRST-exact expression\footnote{This is best seen from the action representation \eqref{g2ff} combined with the already mentioned rescaling $\beta^*\to\frac{\beta^*}{\hat B}$.} and even if it would be, the BRST symmetry is broken, so the standard argument that the expectation value of BRST-exact operators nullifies would anyhow no longer apply\footnote{In the proceeding \cite{Maas:2010wb}, another possible action was formulated to implement the Landau $B$-gauge, but no further details have been provided so far. The proposal of \cite{Maas:2010wb} as it stands is also not respecting BRST.}.

As the BRST breaking is proportional to a lower dimensional operator, it is a soft breaking and hence controllable at the UV level. This is at the core of the renormalization properties of the GZ-like theories. The precise physical consequences of such BRST breaking are not fully understood yet \cite{Brambilla:2014jmp}, despite a series of various recent efforts, \cite{brst1,brst2,brst3,brst4b,brst4,brst5}. Here, we only wish to point out that in a recent numerical effort, further lattice backup have been provided in favour of the BRST breaking by a restriction to the first Gribov region \cite{Cucchieri:2014via}. More work is underway concerning the search for a (lattice verifiable) signal of the BRST breaking \cite{Capri:2014bsa,brst6}.

\end{document}